\documentclass[conference]{IEEEtran}
\IEEEoverridecommandlockouts

\usepackage{cite}
\usepackage{amsmath,amssymb,amsfonts,bm,amsthm}
\usepackage{algorithm}
\usepackage{algorithmic}
\usepackage{array}
\usepackage{multirow}
\usepackage[caption=false,font=footnotesize]{subfig}
\usepackage{textcomp}
\usepackage{stfloats}
\usepackage{url}
\usepackage{graphicx}
\usepackage{booktabs}
\usepackage{xcolor}
\usepackage{balance}
\usepackage{enumitem}
\allowdisplaybreaks[4]

\newcommand{\R}{\mathbb{R}}
\newcommand{\E}{\mathbb{E}}
\newcommand{\cL}{\mathcal{L}}
\newcommand{\cR}{\mathcal{R}}

\newcommand{\sigmoid}{\mathrm{sigmoid}}

\begin{document}

\title{
    Domain-Generalized Adaptive Semantic Communication for Collaborative Perception
    \thanks{This work was supported in part by the National Natural
    Science Foundation of China under Grant No.~62595731 and Grant No.~U22B2001;
    in part by the Program of Jiangsu Province under Grant NTACT-2024-Z-001.
    }
}

\author{
  \IEEEauthorblockN{Fan Gao$^{1,2}$, Youzheng Wang$^{1,2}$, and Ning Ge$^{2}$}
  \IEEEauthorblockA{$^1$Department of Electronic Engineering, BNRist, Tsinghua University, Beijing, China \\
    $^2$Department of Electronic Engineering, Tsinghua University, Beijing, China \\
    gaof23@mails.tsinghua.edu.cn, \{yzhwang, gening\}@tsinghua.edu.cn}
}

\maketitle

\begin{abstract}
We propose RSTA, a domain-generalized semantic communication framework enabling source-free V2X collaborative perception under both observation-domain shift and unseen wireless channel conditions. In V2X, received semantic tokens suffer coupled degradation from pre-transmission domain drift and in-transit channel corruption; existing methods address only one source, leaving adaptation misled by tokens that are simultaneously off-domain and physically degraded. RSTA trains a pre-deployment semantic encoder for transmission stability via cross-domain prototype alignment and cross-channel gradient consistency, and updates a lightweight in-deployment decoder adapter through reliability-gated entropy minimization that restricts gradients to tokens ranked high in both semantic relevance and channel fidelity. A theoretical task robustness decomposition links each loss term to a distinct degradation source, grounding each algorithmic component in a measurable error mode. Trained on AWGN and tested on unseen Rayleigh fading, RSTA achieves +7.2 AP@0.7 over pre-deployment domain generalization on cross-weather tasks and +5.5 on cross-dataset tasks across four V2X benchmarks, updating only 0.21\% of parameters in-deployment with zero inter-agent synchronization overhead.
\end{abstract}

\begin{IEEEkeywords}
Semantic communication, collaborative perception, V2X, domain generalization, deployment-time adaptation.
\end{IEEEkeywords}

\section{Introduction}
Task-oriented semantic communication studies how to encode and transmit only task-relevant information under strict bandwidth and channel constraints~\cite{Xie2021DeepSC,bourtsoulatze2019deepjscc,Hu2022Where2comm,Shao2024TOCOMV2I}. Collaborative perception in V2X is a representative downstream task for this paradigm: connected vehicles exchange compact features to compensate for occlusion and limited sensing range~\cite{Liu2020When2com,Xu2022OPV2V,Zhou2025V2XPnP}. Since transmitting raw sensor streams or full BEV features is infeasible on practical links, the quality of semantic communication directly determines final detection accuracy. Recent work therefore adopts learned semantic coding and reports clear gains in the perception--bandwidth trade-off~\cite{Xu2023SemanticCommunication,Gan2026SComCP,Lu2025CMSC,Gan2025CoDS}.

In deployment, task-oriented semantic communication for collaborative perception faces two coupled shifts. On the observation side, weather, scene layout, and dataset changes alter feature distributions before encoding. On the transmission side, dynamic SNR and fading perturb encoded features in transit. Because the receiver only sees tokens after the channel, semantic mismatch and channel corruption are entangled in the same received features.

Existing anti-shift methods only partially fit this setting. Domain generalization learns cross-domain invariance~\cite{Lin2022BIRM,Rame2022Fishr,Cha2021SWAD,Li2022IIB} and has been extended to collaborative perception~\cite{Li2025V2XDG,Li2025V2XDGW}, but usually ignores distortion induced by the channel. Test-time adaptation and source-free transfer~\cite{Wang2021Tent,Niu2022EATA,Shin2024LTTA,Liang2020SHOT} adapt at deployment without labels, but are mostly designed for clean inputs with domain shift rather than tokens corrupted during communication. Methods in task-oriented communication such as~\cite{Li2025DistShift} target single-user classification, while collaborative perception methods such as CoPEFT~\cite{Wei2025CoPEFT} require labeled target data. A method tailored to task-oriented semantic communication for collaborative perception must handle observation shift, channel shift, and source-free adaptation simultaneously.

To this end, we propose RSTA, a reliability-gated in-deployment adaptation framework for domain-generalized collaborative semantic communication. Pre-deployment training learns semantic representations that remain stable across source domains and channel conditions. In-deployment adaptation updates only a lightweight receiver adapter and gates adaptation by the semantic relevance, channel fidelity, and predictive uncertainty of each token, so unreliable tokens do not dominate gradient updates. Our contributions are summarized as follows:
\begin{itemize}[leftmargin=1.2em]
    \item We propose RSTA, a robust V2X semantic communication framework for collaborative perception, comprising a pre-deployment domain-generalized semantic encoder for transmission stability and an in-deployment source-free receiver adapter updated by reliability-gated entropy minimization.
    \item We derive a task robustness bound via risk decomposition that attributes detection degradation under joint observation--transmission shift to four measurable terms (prototype drift, code energy, channel sensitivity, and token entropy), establishing a principled correspondence between each loss component and a distinct degradation source in task-oriented semantic communication.
    \item Experiments on four benchmarks show $+7.2$ AP@0.7 over pre-deployment DG under weather shift and $+5.5$ under data shift, with $0.21\%$ in-deployment parameter updates and zero inter-agent synchronization overhead.
\end{itemize}

\section{System Model and Problem Formulation}

\begin{figure*}[t]
    \centering
    \includegraphics[width=\linewidth]{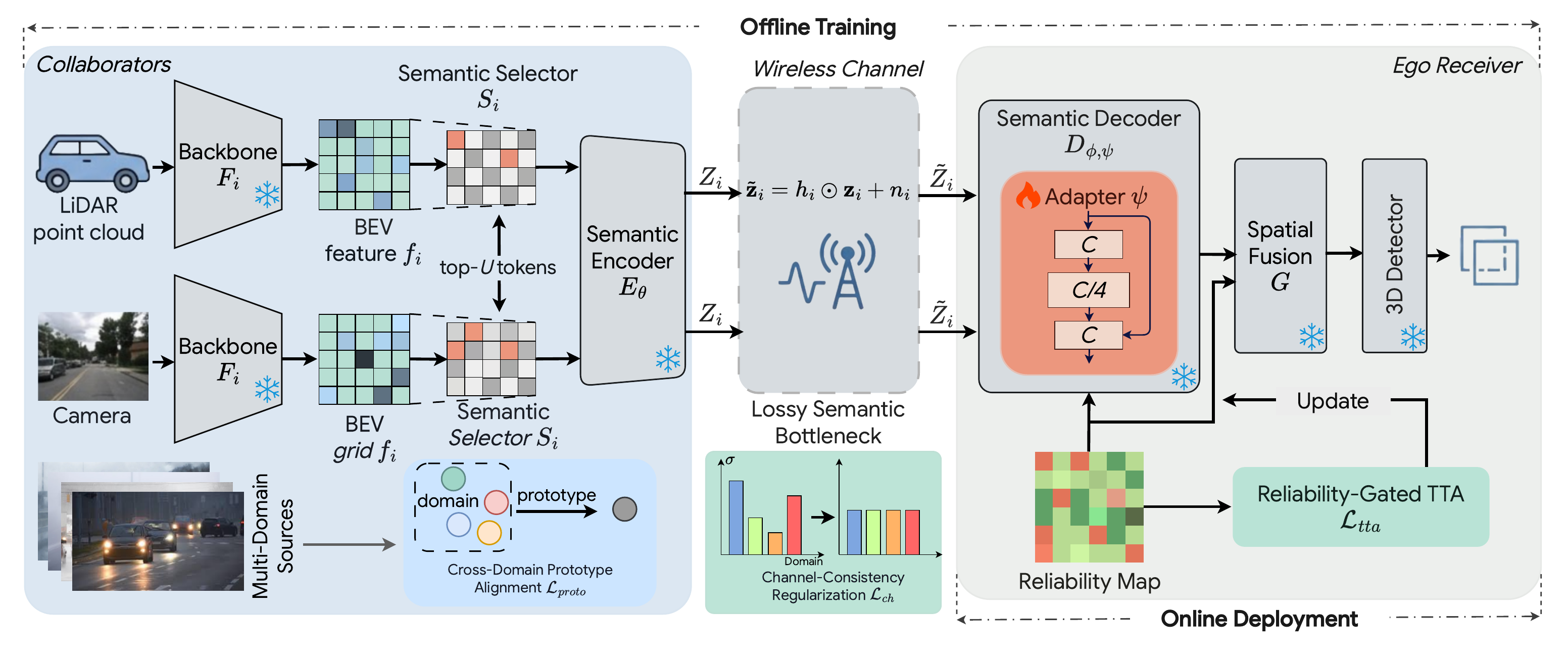}
    \caption{Task-oriented semantic communication for collaborative perception. Each collaborator extracts task-relevant features, maps them into a compact payload, transmits them over a non-ideal V2X link, and the ego receiver performs reliability-aware semantic decoding and fusion.}
    \label{fig:framework}
\end{figure*}

The system in Fig.~\ref{fig:framework} implements task-oriented semantic communication for one ego vehicle and $N$ collaborators. For collaborator $i$, the local observation $x_i$ is first converted into a BEV feature map
\begin{equation}
f_i = F_i(x_i) \in \R^{H\times W\times C},
\end{equation}
where $F_i(\cdot)$ denotes the local perception backbone. Since transmitting the full feature map is prohibitively expensive over the radio, a learned semantic selector $S_i(\cdot)$ keeps only the top-$U$ critical tokens under a payload budget, and a shared semantic encoder maps them into a compact codeword tensor
\begin{equation}
z_i = E_\theta\big(S_i(F_i(x_i))\big) \in \R^{U\times d},
\end{equation}
where $d$ is the code dimension per token. These semantic symbols are then delivered over a V2X link. Under a channel coefficient $h_i$ and additive noise $n_i$, the ego receives
\begin{equation}
\tilde z_i = h_i \odot z_i + n_i,
\label{eq:channel}
\end{equation}
where $\odot$ denotes element-wise multiplication. This per-symbol model is standard in deep JSCC~\cite{bourtsoulatze2019deepjscc,Xie2021DeepSC}; it captures flat fading, SNR fluctuation, and equivalent noise from quantization or packet loss.

The ego receiver applies a semantic decoder $D_{\phi,\psi}$ to each received code, fuses the decoded multi-agent features through a spatial fusion module $G(\cdot)$, and predicts 3D objects using a detection head $C(\cdot)$:
\begin{equation}
\hat y = C\Big(G\big(D_{\phi,\psi}(\tilde z_0),\ldots,D_{\phi,\psi}(\tilde z_N)\big)\Big),
\end{equation}
where $\phi$ collects the main decoder parameters and $\psi$ denotes a lightweight trainable adapter inserted before fusion. In our implementation, the adapter is a two-layer bottleneck MLP with a residual connection, mapping $C\!\to\!C/4\!\to\!C$.

Deployment introduces two coupled mismatches. \emph{Observation-side shift} occurs when weather, road layout, or dataset change alter the feature distribution before encoding. \emph{Transmission-side shift} occurs when dynamic channels and payload distortion perturb the code in transit. The receiver observes only the code $\tilde z_i$ after the channel, so both effects are folded into the same token, defining a \emph{joint observation--transmission shift} problem.

In the pre-deployment stage, we have $M$ labeled source domains $\{\mathcal D_m\}_{m=1}^M$ and $K$ training channel distributions $\{\nu_k\}_{k=1}^K$. At deployment, both the target domain and test channel may be unseen, and source data are unavailable. The goal is to preserve communication reliability for downstream perception while confining in-deployment adaptation to a tiny on-device module:
\begin{equation}
\begin{split}
    \min_{\theta,\phi}\; \cR_s & \quad \text{pre-deployment}, \\
    \min_{\psi}\; \cR_t & \quad \text{in-deployment with } |\psi|\ll |\theta|+|\phi|,
\end{split}
\end{equation}
where $\cR_s$ and $\cR_t$ denote the expected detection loss on source and target distributions, respectively. The constraint $|\psi|\ll|\theta|+|\phi|$ enforces edge-feasible in-deployment adaptation without updating the full semantic pipeline.

\section{Proposed Method}
\subsection{Pre-Deployment Transmission-Stable Semantic Learning}
The pre-deployment stage learns semantic codes that remain useful for the downstream task after both compression and wireless delivery. Unlike conventional DG, all regularization applies to decoded features after the wireless channel, since that is what the ego receiver observes.

Let $f_u$ denote the decoded feature of token $u$, $q_{u,c}$ its soft class assignment, and $C_{\mathrm{cls}}$ the number of semantic classes. For source domain $m$, the class-$c$ decoded prototype is
\begin{equation}
\mu_{m,c}=\frac{\sum_{u\in\mathcal D_m} q_{u,c}f_u}{\sum_{u\in\mathcal D_m} q_{u,c}+\varepsilon}, \qquad
\bar\mu_c=\frac{1}{M}\sum_{m=1}^M \mu_{m,c}.
\label{eq:anchoring}
\end{equation}
We reduce cross-domain semantic drift by minimizing
\begin{equation}
\cL_{\mathrm{proto}}=\frac{1}{C_{\mathrm{cls}}}\sum_{c=1}^{C_{\mathrm{cls}}}\frac{1}{M}\sum_{m=1}^M\|\mu_{m,c}-\bar\mu_c\|_2^2.
\label{eq:lproto}
\end{equation}
This term aligns the semantic geometry at the receiver across source domains, reducing sensitivity to observation shift.

To improve robustness to channel variation, we regularize the sensitivity of the semantic decoder across training channels. For channel $k$, let $\ell_k$ be the detection loss and $G_k^\phi\in\R^{|\phi|}$ the diagonal of the per-sample gradient covariance $\mathrm{diag}\!\big(\mathrm{Cov}_{x\sim\mathcal D_k}[\nabla_\phi \ell_k(x)]\big)$, following the Fishr principle~\cite{Rame2022Fishr}. We define a cross-channel gradient consistency penalty
\begin{equation}
\cL_{\mathrm{ch}}=\frac{1}{K}\sum_{k=1}^K \left\|G_k^\phi-\frac{1}{K}\sum_{j=1}^K G_j^\phi\right\|_1.
\label{eq:lch}
\end{equation}
Matching these gradient statistics discourages semantic features from over-specializing to one training channel and stabilizes the decoder under unseen fading laws.

Communication reliability in task-oriented semantic coding also depends on code energy. High-energy codes are more sensitive to channel variation and require larger radio resources. We therefore add a compactness term
\begin{equation}
\cL_{\mathrm{cmp}}=\frac{1}{NUd}\sum_{i=1}^N \|z_i\|_F^2,
\label{eq:lcmp}
\end{equation}
which acts both as a transmit power surrogate and as a regularizer on channel sensitivity.

Finally, we pretrain a lightweight uncertainty head so that the receiver can later judge whether a token should contribute to in-deployment adaptation. For two independently sampled training channels, let $p_u^{(1)}$ and $p_u^{(2)}$ be the corresponding posteriors. Their symmetric disagreement is
\begin{equation}
d_u = \tfrac12\Big(\mathrm{KL}(p_u^{(1)}\Vert p_u^{(2)})+\mathrm{KL}(p_u^{(2)}\Vert p_u^{(1)})\Big),
\end{equation}
and we regress the uncertainty estimate through
\begin{equation}
\cL_{\mathrm{unc}}=\frac{1}{U}\sum_u |\hat\sigma_u-d_u|.
\label{eq:lunc}
\end{equation}
The combined pre-deployment task loss is
\begin{equation}
\cL_{\mathrm{off}}=\cL_{\mathrm{det}}+\lambda_p\cL_{\mathrm{proto}}+\lambda_c\cL_{\mathrm{ch}}+\lambda_n\cL_{\mathrm{cmp}}+\lambda_u\cL_{\mathrm{unc}}.
\label{eq:loff}
\end{equation}

\subsection{Reliability-Gated Receiver for Adaptation}
At deployment, only unlabeled target domain streams are available. We adapt a lightweight adapter at the receiver rather than the sender. Broadcasting updated encoder weights would consume control plane bandwidth, and independently adapted senders risk drifting into incompatible code spaces. Receiver adaptation avoids both problems: zero signaling between agents is needed, and shared semantic features remain aligned across collaborators.

Not every received token should contribute to adaptation: a token may carry high task relevance but low physical reliability, or vice versa. We define a reliability score
\begin{equation}
r_u=\sigmoid\big(w_1 a_u+w_2\rho_u-w_3\hat\sigma_u\big)\in[0,1],
\label{eq:rel}
\end{equation}
where $a_u$ is the semantic importance score from the sender, $\rho_u$ is a channel confidence statistic at the receiver derived from the received signal magnitude, and $\hat\sigma_u$ is the predicted token uncertainty. The weights $w_1,w_2,w_3$ are learned jointly in pre-deployment training. A large $r_u$ means that the token is both semantically useful and physically trustworthy.

\begin{figure}[t]
    \centering
    \includegraphics[width=\linewidth]{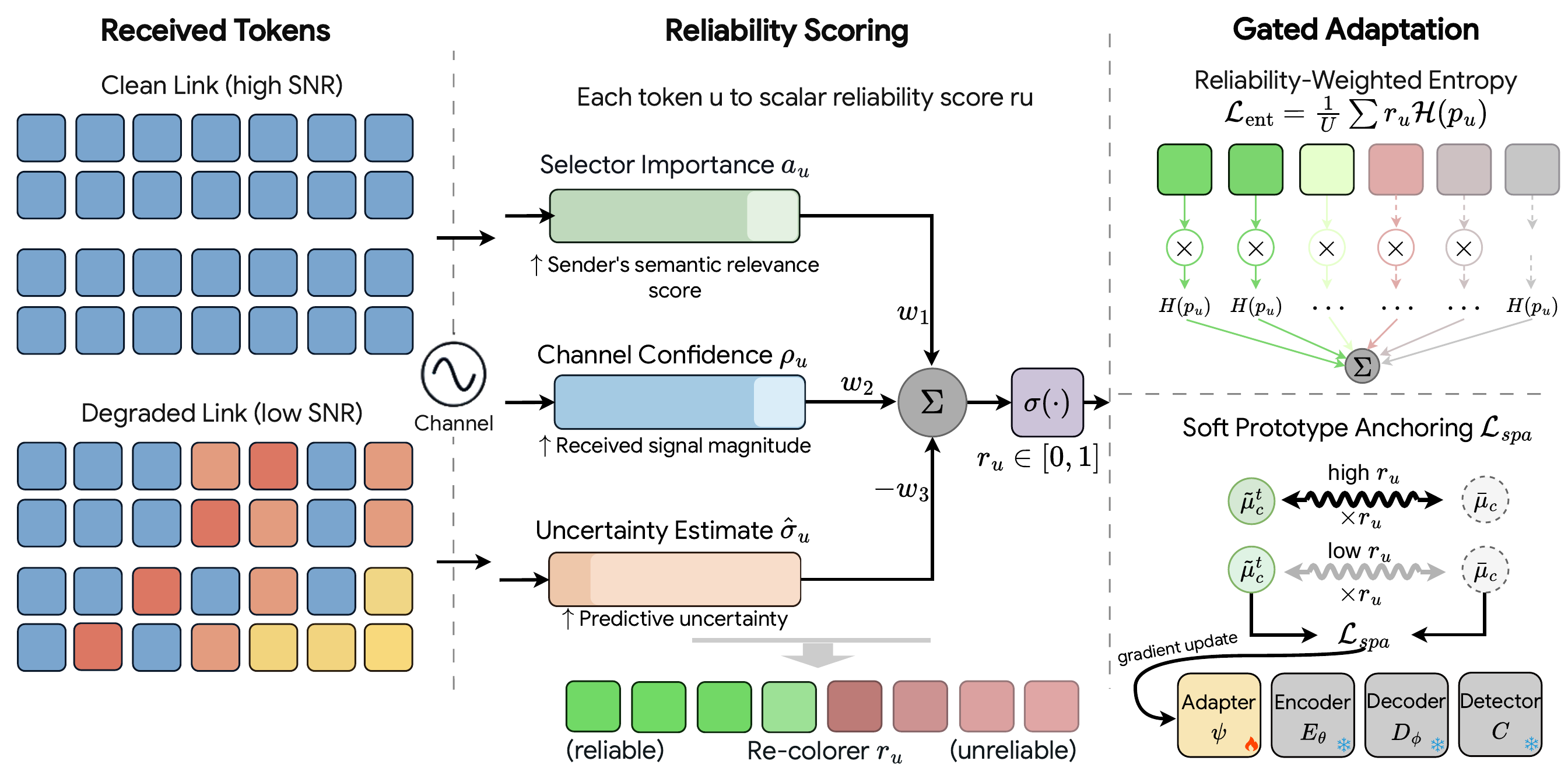}
    \caption{Source-free in-deployment adaptation with reliability gating. The ego receiver decodes incoming communication codes, estimates token reliability from semantic relevance and channel confidence, forms entropy and prototype alignment losses from reliable target features, and updates only the lightweight adapter.}
    \label{fig:tta}
\end{figure}

Online adaptation is then driven by reliability-gated entropy minimization,
\begin{equation}
\cL_{\mathrm{ent}}=\frac{1}{U}\sum_{u=1}^{U} r_u\,\mathcal H(p_u),
\label{eq:lent}
\end{equation}
so unreliable tokens contribute little to the gradient. To prevent the target feature space from drifting away from the semantic manifold learned from source data, we also impose a soft prototype anchoring term
\begin{equation}
\tilde\mu_c^t=\frac{\sum_{u} r_u p_{u,c} f_u}{\sum_u r_u p_{u,c}+\varepsilon}, \quad
\cL_{\mathrm{spa}}=\frac{1}{C_{\mathrm{cls}}}\sum_c\|\tilde\mu_c^t-\bar\mu_c\|_2^2.
\label{eq:lspa}
\end{equation}
Only the adapter parameters $\psi$ are updated,
\begin{equation}
\cL_{\mathrm{tta}}=\cL_{\mathrm{ent}}+\lambda_s\cL_{\mathrm{spa}}+\lambda_r\|\psi-\psi_0\|_2^2,
\label{eq:ltta}
\end{equation}
where the last term keeps the in-deployment model close to the source checkpoint $\psi_0$. Fig.~\ref{fig:tta} summarizes this source-free adaptation loop.

\subsection{Task Robustness Analysis}
The task detection risk on the deployment domain admits a communication-oriented decomposition that reveals the function of each loss term. Assume the detection loss $\ell$ is $L_f$-Lipschitz in the decoded feature and $L_d$-L
ipschitz in the received code. Partition the token set into reliable ($r_u\!\ge\!\tau$) and unreliable ($r_u\!<\!\tau$) subsets. The target risk decomposes (up to constants) as
\begin{equation}
\begin{aligned}
\cR_t \lesssim\;&
\underbrace{L_f\,\Delta_{\mathrm{proto}}}_{\text{prototype drift}}
+ \underbrace{L_d\,\E\|z_i\|_2\,\E\|\Delta h\|_2}_{\text{channel sensitivity}} \\
&+ \underbrace{\E_t[1-r_u]\,\ell_{\max}}_{\text{unreliable mass}}
+ \underbrace{\E_t[r_u\,\mathcal H(p_u)]}_{\text{reliable entropy}} .
\end{aligned}
\label{eq:risk}
\end{equation}
where $\Delta_{\mathrm{proto}}=\frac{1}{C_{\mathrm{cls}}}\sum_c\|\tilde\mu_c^t-\bar\mu_c\|_2^2$, $\Delta h=h-h'$ denotes channel mismatch, and $\ell_{\max}$ is the worst-case per-token loss.

\begin{itemize}[leftmargin=1.2em,itemsep=2pt]
    \item \textbf{Semantic prototype drift.}
    The first term is the squared distance between target and source prototypes, directly minimized by $\cL_{\mathrm{spa}}$ in deployment.

    \item \textbf{Channel sensitivity.}
    Under \eqref{eq:channel}, $\|(h{-}h')\odot z_i\|_2\leq\|z_i\|_2\|h{-}h'\|_2$; reducing $\E\|z_i\|_2$ via $\cL_{\mathrm{cmp}}$ tightens this bound and limits transmit energy.

    \item \textbf{Unreliable token mass.}
    Reliability gating prevents low-$r_u$ tokens from dominating adaptation; $\cL_{\mathrm{ch}}$ and $\cL_{\mathrm{unc}}$ reduce the unreliable mass in pre-deployment training.

    \item \textbf{Reliable token entropy.}
    Residual prediction uncertainty among reliable tokens is directly reduced by $\cL_{\mathrm{ent}}$.
\end{itemize}

\noindent Plain entropy minimization~\cite{Wang2021Tent} reduces only the last term and treats all tokens identically regardless of whether degradation is due to domain mismatch or channel corruption; that conflation accounts for its weaker performance under joint shift.

\begin{algorithm}[t]
\caption{In-Deployment Reliability-Gated Receiver Adaptation}
\label{alg:rtsa}
\begin{algorithmic}[1]
\STATE Load pre-deployment model $\{\theta,\phi,\psi_0\}$ and source anchors $\{\bar\mu_c\}$
\FOR{each unlabeled target batch}
    \STATE Receive communication codes $\{\tilde z_i\}$ and decode features $\{f_u\}$
    \STATE Compute posteriors $\{p_u\}$ and reliability scores $\{r_u\}$ by \eqref{eq:rel}
    \STATE Form $\cL_{\mathrm{ent}}$ by \eqref{eq:lent} and $\cL_{\mathrm{spa}}$ by \eqref{eq:lspa}
    \STATE Update only adapter parameters $\psi$ with one or two SGD/Adam steps on \eqref{eq:ltta}
    \STATE Run adapted inference for the current batch
\ENDFOR
\end{algorithmic}
\end{algorithm}

\section{Experiments}
\subsection{Simulation Settings}
We evaluate on four V2X collaborative perception datasets. OPV2V~\cite{Xu2022OPV2V}, V2XSet~\cite{Xu2022V2XViT}, V2V4Real~\cite{Xu2023V2V4Real}, and DAIR-V2X~\cite{Yu2022DAIR}. We run two deployment protocols. In the \emph{weather-shift} setting, models trained on clean OPV2V and V2XSet are tested on OPV2V-w and V2XSet-w with fog, rain, and snow~\cite{Li2025V2XDGW}. In the \emph{data-shift} setting, the source is OPV2V and V2XSet jointly while the target is V2V4Real or DAIR-V2X, with results averaged across target sets. Training uses AWGN with SNR $\in[0,18]$~dB; test channels include matched AWGN and unseen Rayleigh fading, creating a realistic joint observation--transmission shift for the receiver.

All experiments were implemented in PyTorch and conducted on a server with four NVIDIA A800 (80\,GB) GPUs. We use a PointPillar-based intermediate-fusion detector with AttnFuse~\cite{Xu2022OPV2V} and a lightweight semantic codec inserted between collaborator BEV features and ego fusion. Each collaborator retains the top-$16$ semantic tokens, and each token is encoded into $64$ real-valued channel symbols, yielding $1024$ channel uses per collaborator per frame and a compression ratio of $128\times$ relative to the full BEV feature map. Pre-deployment training uses AdamW with learning rate $2\times10^{-4}$ for $24$ epochs. The in-deployment adapter is optimized with AdamW at learning rate $2\times10^{-5}$ for two steps per target batch. Unless otherwise stated, $\lambda_p=0.2$, $\lambda_c=0.1$, $\lambda_n=5\times10^{-4}$, $\lambda_u=0.5$, $\lambda_s=0.3$, and $\lambda_r=10^{-4}$.

\begin{table}[t]
    \centering
    \caption{Comparison between receiver and sender adaptation.}
    \label{tab:scheme}
    \footnotesize
    \resizebox{\linewidth}{!}{
        \begin{tabular}{lcccc}
        \toprule
        Scheme & AP@0.7 & Updated params & Ctrl. overhead & Latency \\
        \midrule
        Pre-Deployment DG & 54.5 & 0\% & 0 kb & 61.9 ms \\
        Sender-side stem & 60.3 & 0.43\% & 14.6 kb/round & 74.8 ms \\
        Receiver-side adapter & \textbf{61.7} & \textbf{0.21\%} & \textbf{0 kb/round} & \textbf{70.8 ms} \\
        \bottomrule
        \end{tabular}
    }
\end{table}

We compare three baselines. \textit{DeepJSCC}~\cite{bourtsoulatze2019deepjscc} trains the semantic codec on a single source domain without any domain generalization or in-deployment adaptation. \textit{Pre-Deployment DG} applies MLDG~\cite{li2018learning} during pre-deployment training. \textit{TENT}~\cite{Wang2021Tent} augments MLDG with entropy minimization that updates all batch normalization affine parameters at deployment time. The \textit{Upper bound} trains and tests on the target domain with matched channel conditions, representing the performance ceiling when no distribution shift exists. All baselines share the same backbone, fusion, selector, and semantic codec; differences isolate how each method handles reliability. Detection quality is measured by Average Precision at IoU thresholds $0.5$ and $0.7$ (AP@0.5, AP@0.7) for weather shift, and at thresholds $0.3$ and $0.5$ (AP@0.3, AP@0.5) for data shift.

\subsection{Main Results Under Joint Observation--Transmission Shift}
\begin{figure}[t]
    \centering
    \includegraphics[width=\linewidth]{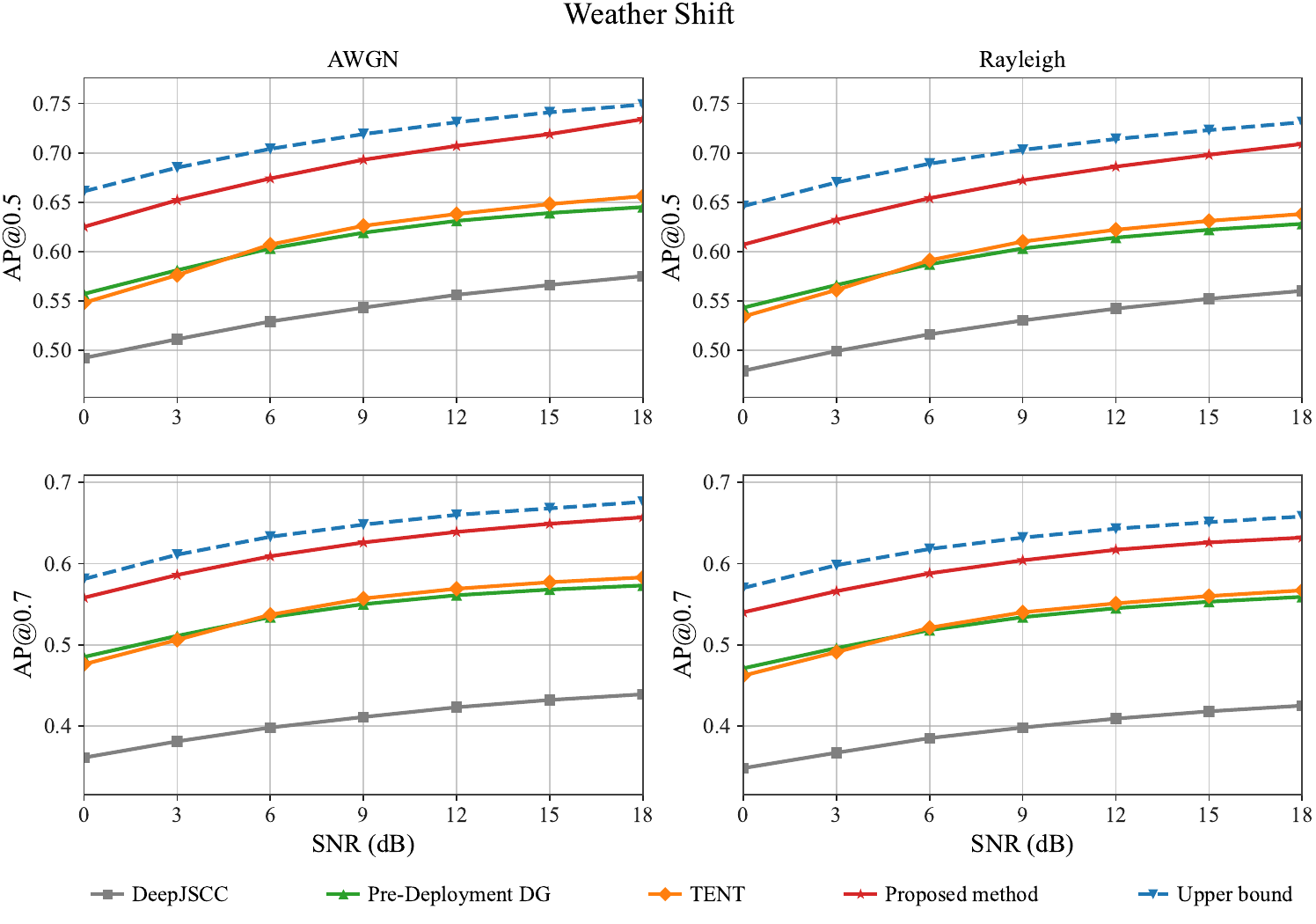}
    \caption{Weather shift under AWGN and Rayleigh test channels. Left: AWGN. Right: Rayleigh. Top: AP@0.5. Bottom: AP@0.7.}
    \label{fig:weather_shift_results}
\end{figure}

\begin{figure}[t]
    \centering
    \includegraphics[width=\linewidth]{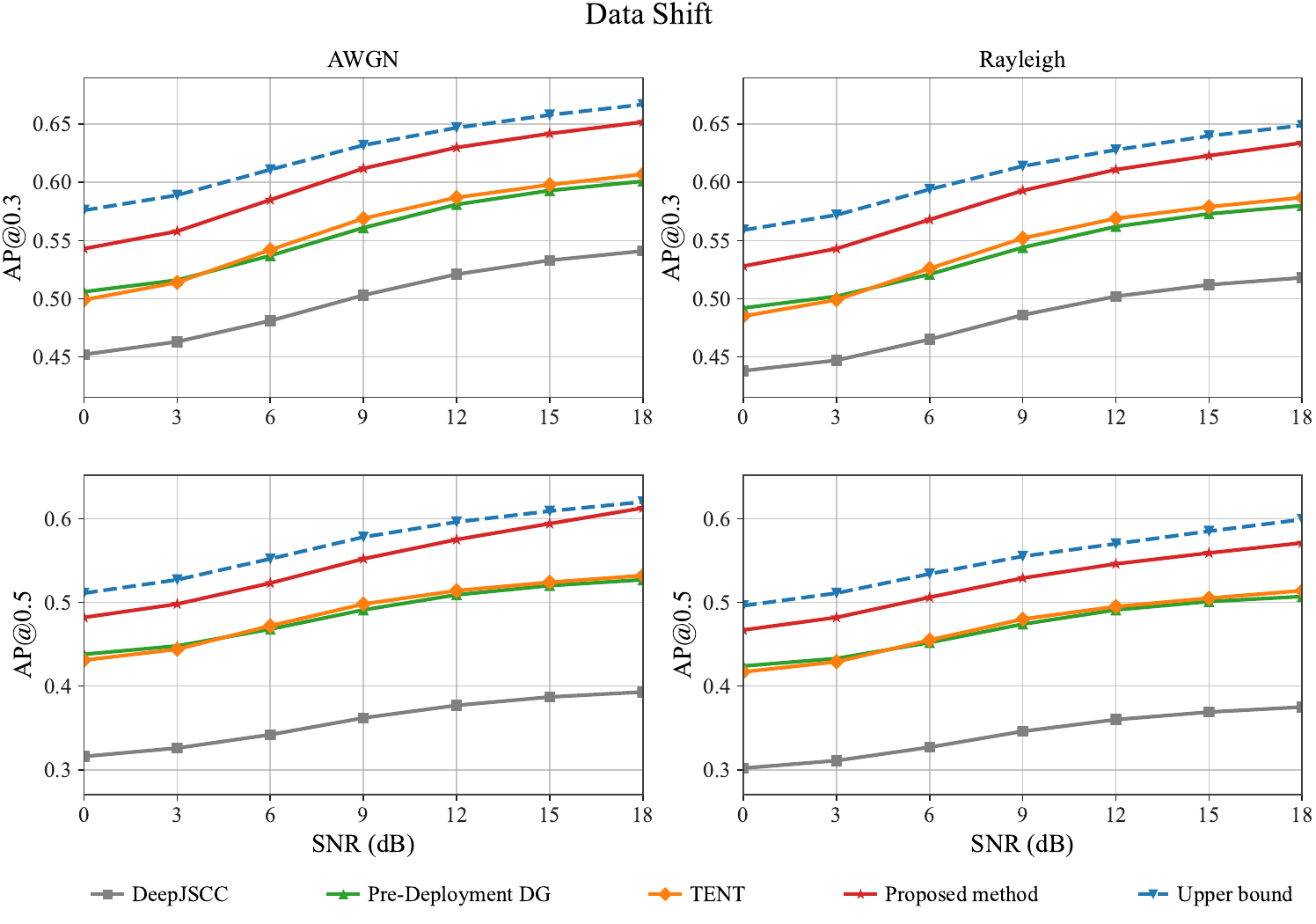}
    \caption{Data shift under AWGN and Rayleigh test channels. Left: AWGN. Right: Rayleigh. Top: AP@0.3. Bottom: AP@0.5.}
    \label{fig:data_shift_results}
\end{figure}

Figs.~\ref{fig:weather_shift_results} and \ref{fig:data_shift_results} plot detection performance versus SNR under weather shift and data shift, respectively. Under AWGN, methods saturate above $12$~dB and the residual gap is dominated by observation mismatch; under Rayleigh fading, DeepJSCC and TENT suffer a pronounced low-SNR drop while RSTA preserves a clear margin by suppressing unreliable tokens. At $12$~dB Rayleigh, RSTA reaches $61.7$ AP@0.7 on weather shift ($+7.2$ over Pre-Deployment DG) and $54.6$ AP@0.5 on data shift ($+5.5$), updating only $0.21\%$ of parameters. DeepJSCC drops to $40.9$ AP@0.7 from coupled domain--channel degradation, while TENT closes only a fraction of the gap because corrupted tokens mislead its uniform update. The small AWGN-to-Rayleigh margin ($63.9\to 61.7$ AP@0.7 on weather shift, $57.5\to 54.6$ AP@0.5 on data shift) confirms that channel confidence $\rho_u$ limits the fading penalty through the reliability gate. Under data shift, the larger improvement on AP@0.5 than AP@0.3 indicates that reliability-gated adaptation is most effective when tight semantic alignment after transmission is needed. Consistent gains under both shift types confirm that the method generalizes across appearance-level and scene-level domain changes. Table~\ref{tab:scheme} compares adaptation placement; adapting at the receiver achieves higher AP@0.7 than adapting at the sender, with zero control overhead.

\subsection{Ablation on Semantic Transport Reliability}
Table~\ref{tab:ablation} isolates each component. Removing $\cL_{\mathrm{ch}}$ produces the largest drop, confirming that channel sensitivity dominates under Rayleigh fading. Replacing reliability gating with uniform weighting ($r_u\!=\!1$) lets corrupted tokens mislead the update. Each ablation traces to a distinct degradation term in \eqref{eq:risk}. Pre-deployment channel consistency and in-deployment gating are not redundant: $\cL_{\mathrm{ch}}$ reduces the number of corrupted tokens below the reliability threshold, while gating suppresses their gradient weight, and removing either degrades performance.

\begin{table}[t]
\centering
\caption{Ablation on the joint-shift setting.}
\label{tab:ablation}
\setlength{\tabcolsep}{3.4pt}
\footnotesize
\begin{tabular}{lccc}
\toprule
Variant & AP@0.5 & AP@0.7 & Interpretation \\
\midrule
w/o $\cL_{\mathrm{proto}}$ & 66.0 & 59.6 & semantic drift \\
w/o $\cL_{\mathrm{ch}}$ & 65.3 & 59.0 & channel sensitivity \\
Uniform ent.\ min.\ ($r_u\!=\!1$) & 65.8 & 59.2 & unreliable tokens \\
w/o $\cL_{\mathrm{spa}}$ & 67.4 & 60.5 & prototype drift \\
Full method & \textbf{68.6} & \textbf{61.7} & -- \\
\bottomrule
\end{tabular}
\end{table}

\subsection{Qualitative Visualization}
\begin{figure*}[t]
    \centering
    \includegraphics[width=\linewidth]{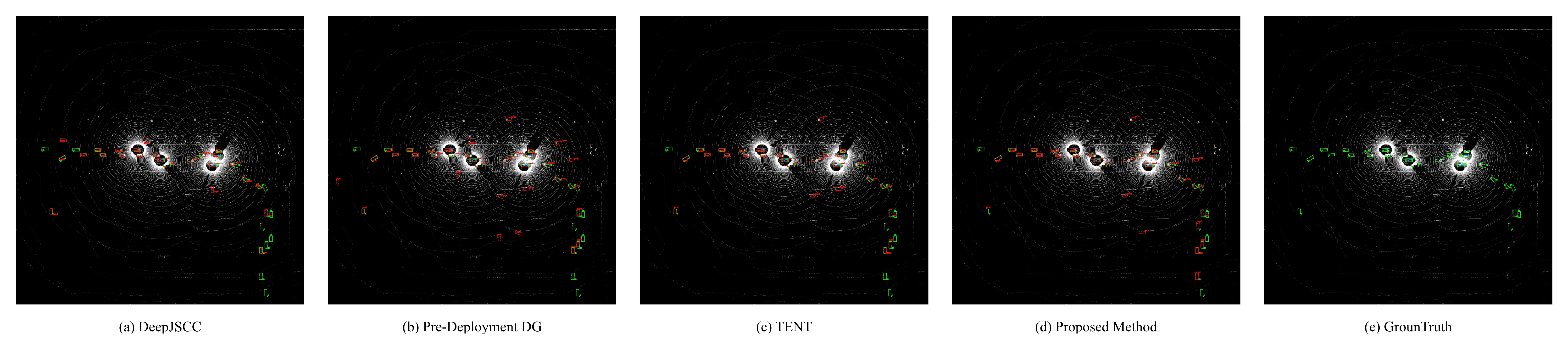}
    \caption{Qualitative BEV detection results under weather shift with Rayleigh fading at SNR${}={}$12~dB. Green boxes denote correct detections; red boxes denote false positives or mislocated objects. The proposed method most closely matches the ground truth by suppressing unreliable tokens during in-deployment adaptation.}
    \label{fig:visual}
\end{figure*}

Fig.~\ref{fig:visual} visualizes BEV detection under weather shift at SNR${}={}$12~dB with Rayleigh fading. DeepJSCC produces scattered false positives due to coupled domain--channel degradation. TENT introduces additional spurious boxes from indiscriminate adaptation on corrupted tokens. RSTA most closely matches the ground truth, confirming that reliability gating yields accurate localization with fewer false positives.

\subsection{Complexity, Latency, and Communication Overhead}
Table~\ref{tab:lat} shows that the in-deployment adapter adds only $8.9$~ms to the $61.9$~ms frozen pipeline, updating $0.07$~M of $33.27$~M total parameters. Reliability gating also provides a fail-safe: at $0$~dB SNR, most scores collapse toward zero so the system reverts to the pre-deployment model, losing less than $0.5$ AP@0.7, whereas uniform entropy minimization degrades by $1.5$.

\begin{table}[t]
\centering
\caption{Latency and overhead.}
\label{tab:lat}
\setlength{\tabcolsep}{3.2pt}
\footnotesize
\begin{tabular}{lccc}
\toprule
Component & Params & Latency & Notes \\
\midrule
Feature extraction & 21.84 M & 31.5 ms & frozen \\
Semantic selector + codec & 1.96 M & 7.9 ms & frozen \\
Receiver decoder/fusion & 9.40 M & 22.5 ms & frozen except adapter \\
In-deployment adapter update & 0.07 M & 8.9 ms & 2 Adam steps \\
Total & 33.27 M & 70.8 ms & $0.21\%$ updated \\
\bottomrule
\end{tabular}
\end{table}

\section{Conclusion}
RSTA improves robustness in task-oriented semantic communication for V2X collaborative perception by treating observation-domain shift and wireless channel distortion as a coupled problem. Pre-deployment training learns domain- and channel-stable semantic features, and source-free in-deployment reliability gating enables edge-feasible adaptation with only $0.21\%$ parameter updates and no inter-agent signaling overhead. Extending RSTA to frequency-selective OFDM channels and adaptive gating thresholds is a natural direction for future work.

\balance
\bibliographystyle{IEEEtran}
\bibliography{IEEEabrv,rsta}

\end{document}